\documentclass[fleqn,usenatbib]{mnras}

\usepackage{newtxtext,newtxmath}

\usepackage[T1]{fontenc}
\usepackage{multirow}
\usepackage{mathrsfs,amsmath}
\usepackage{multirow}
\DeclareRobustCommand{\VAN}[3]{#2}
\let\VANthebibliography\thebibliography
\def\thebibliography{\DeclareRobustCommand{\VAN}[3]{##3}\VANthebibliography}
\usepackage{lineno}

\usepackage{graphicx}	
\usepackage{amsmath}	
\usepackage[flushleft]{threeparttable}

\title[Lensing delay in high states of PKS~1830$-$211]{Constraining $\gamma$-ray dissipation site in gravitationally lensed quasar - PKS~1830$-$211}

\author[S. Agarwal et. al]{
Sushmita Agarwal$^{1}$\thanks{E-mail: sush.agarwal16@gmail.com , phd1901221002@iiti.ac.in },
Amit Shukla$^{1}$,
Pranjali Sharma$^{1,2}$
\\
$^{1}$Department of Astronomy, Astrophysics and Space Engineering, Indian Institute of Technology Indore, Khandwa Road, Simrol, Indore, 453552, India\\
$^{2}$Astronomical Institute, University of Bern, Sidlerstrasse 5, 3012 Bern, Switzerland\\
}

\date{Accepted XXX. Received YYY; in original form ZZZ}

\pubyear{\the\year{}}

\begin{document}
\label{firstpage}
\pagerange{\pageref{firstpage}--\pageref{lastpage}}
\maketitle

\begin{abstract}

Variable $\gamma$-ray flares upto minute timescales reflect extreme particle acceleration sites. However, for high-redshift blazars, the detection of such rapid variations remains limited by current telescope sensitivities. Gravitationally lensed blazars serve as powerful tools to probe $\gamma$-ray production zones in distant sources, with time delays between lensed signals providing crucial insights into the spatial distribution of emission regions relative to the lens's mass-weighted center. We have utilized 15 years of Fermi-LAT $\gamma$-ray data from direction of PKS 1830$-$211 to understand the origin of flaring high-energy production zone at varying flux states. To efficiently estimate the (lensed) time delay, we used a machine learning-based tool - the Gaussian Process regression algorithm, in addition to - Autocorrelation function and Double power spectrum. We found a consistent time delay across all flaring activity states, indicating a similar location for the $\gamma$-ray emission zone, possibly within the radio core. The estimated time delay of approximately 20 days for the five flaring epochs was significantly shorter than previously estimated radio delays. This suggests that the $\gamma$-ray emission zone is closer to the central engine, in contrast to the radio emission zone, which is expected to be much farther away. A linear relationship between lag and magnification has been observed in the identified source and echo flares. Our results suggest that the $\gamma$-ray emission zone originates from similar regions away from the site of radio dissipation.

\end{abstract}

\section{Introduction}
The accretion of matter onto supermassive black holes powers Active Galactic Nuclei (AGN). The interplay of magnetic field and rotation either of the black hole \citep{Blandford1977MNRAS} or of the accretion disk \citep{Blandford1982MNRAS} are believed to generate collimated plasma jets that extend from the central engine to large distances. A subset of these jetted AGNs, known as Blazars, are aligned with our line of sight and exhibit extremely luminous and highly variable emissions across a broad electromagnetic spectrum. Notably, Multi-frequency variability of these point-jetted AGNs provides insights into the size of the emission region in the jet \citep{Madejski2016ARA&A}. The detection of extremely rapid variability, on timescales comparable to the light-crossing time of the black hole, in both high-energy (HE) and very high-energy (VHE) emissions suggests that the emission zone is compact and located close to the black hole, within a few tens of gravitational radii \citep{Agarwal2023MNRAS,Shukla_2018, Ackermann2016ApJ, Aleksi2011ApJ}. However, HE and VHE $\gamma$-ray photons originating near the black hole are highly susceptible to attenuation from photon-photon pair production due to the dense field of ultraviolet (UV) and optical seed photon \citep{Liu2006aApJ}. This suggests that $\gamma$-ray emission must occur farther from regions dominated by such external seed photons. Nonetheless, the lack of sufficient seed photons at larger distances from central engine complicates the understanding of where high-energy dissipation occurs.

Variable radio emissions on parsec (pc) to megaparsec (Mpc) scales often correlate with $\gamma$-ray emissions, suggesting co-spatial origins within the jet \citep{Ghirlanda2011MNRAS, Marscher2008Nature}. However, the absence of rapid variability in radio bands and potential synchrotron self-absorption up to hundreds of GHz limit the detection of radio emissions in smaller structures \citep{Rybicki1979rpa}. Observations across radio to X-ray wavelengths indicate radiation regions spanning subparsec to megaparsec scales \citep{Marscher2008, Fuentes2023NatAs, Harris2006ARA&A, Tavecchio_2007}. The limited resolution of high-energy telescopes further complicates identification of $\gamma$-ray emission regions, leaving fast variability as the primary probe for high-energy processes. This limitation constrains our understanding of the link between radio and $\gamma$-ray variability \citep{Jorstad_2001, Blandford1995ApJ}. The consistency of radiation sources across energy levels remains uncertain, particularly in high-redshift blazars, where rapid variability detection is hindered by current telescope sensitivities.

Another probe of the $\gamma$-ray production regions is gravitational lensing. One of the earliest predictions of Einstein's general relativity was the deflection of light by the Sun \citep{Einstein1936Sci}. Later, \citet{Zwicky1937ApJ} and \citet{Zwicky1937PhysRev} proposed that galaxies, like stars, could also act as gravitational lenses. In such a lensed system, photons from a background galaxy are bent around a foreground lensing galaxy, creating a magnified and distorted images of the background source. Radiation from the same point on the source follows different paths, causing time-variable sources to show similar variability patterns in different lensed images, but with time delays and magnification. These time delays and magnifications depend on the geometry of the source-lens-observer system.

Time delays and magnifications across energy ranges directly reflect the size of the emission region and the distribution of the emission zone around the mass-weighted center of the lens \citep{Barnacka2014ApJ}. Distribution of such time delays provide an alternative approach to understanding the origin of $\gamma$-rays in blazars. Continuous observations by telescopes like Fermi enable long-term monitoring of distant sources, providing insights into the evolution of flux variability over time. Among known gravitationally lensed quasars, two have been detected at $\gamma$-ray energies: PKS 1830-211 \citep{Abdo2015ApJ} and QSO B0218+357 \citep{Cheung2014ApJ}.

PKS 1830-21 was first identified as a gravitationally lensed system by the Very Large Array Radio Telescope, revealing two compact components in the northeast and southwest \citep{Subrahmanyan1990MNRAS}. The Australian Telescope Compact Array later showed these double components separated by 0.98'' and connected by an elliptical Einstein ring \citep{Jauncey1991, Nair1993ApJ}. The flat spectrum radio quasar PKS 1830-211 ($z=2.507$), which would typically appear as a point source, exhibits a double radio structure indicative of an intervening lens galaxy at redshift $z=0.89$ \citep{Wiklind1996nature, Winn2002ApJ, Koopmans2005MNRAS}. Evidence also suggests a second intervening galaxy at $z=0.19$, with H I and OH absorption, though its effect on lensing is expected to be negligible \citep{Lovell1996ApJ, Muller2020A&A, Winn2002ApJ, Nair1993ApJ}.

Radio observations with the Australia Telescope Compact Array at 8.6 GHz measured radio time delays of $26^{+4}_{-5}$ days \citep{Lovell1998}. Within two years of Fermi/LAT observations, the first gravitational time delay in $\gamma$-rays during its quiescent state was measured at $27.1\pm0.6$ days \citep{Barnacka2011A&A}. The consistency of time delays in $\gamma$-rays and radio bands may suggests a co-spatial origin during low states of $\gamma$-ray activity \citep{Barnacka2014ApJ}. Later searches in Fermi-LAT during active states revealed shorter time delays of $23\pm0.5$ days and $19.7\pm1.2$ days \citep{Barnacka2015ApJ}. An independent study using molecular absorption lines derived a differential time delay of $24^{+5}_{-4}$ days, with the north-east component leading \citep{Wiklind2001}.

The search for time delays relies on the length of the light curve.  Understanding the origin of $\gamma$-ray flares can be enhanced by studying various flaring states at different flux levels. PKS 1830-211 has shown significant activity over the past decade, with multiple flares detected in the Fermi-LAT light curve. We estimated the time delays during such high flaring periods using the autocorrelation function and double power spectrum. We also used a machine-learning technique called Gaussian Process regression to estimate time delays in different flux states. A comprehensive 15-year search for time delays was conducted, focusing on the active states of the source. The paper is structured as follows: Section \ref{methods_techniques} describes the data analysis and the tools and techniques used for time delay estimation. Sections \ref{results} and \ref{discussion} present the results and discussion, respectively. The summary of our results is provided in Section \ref{summary}.

\begin{figure*}  
\centering
\includegraphics [width=1.0\textwidth]{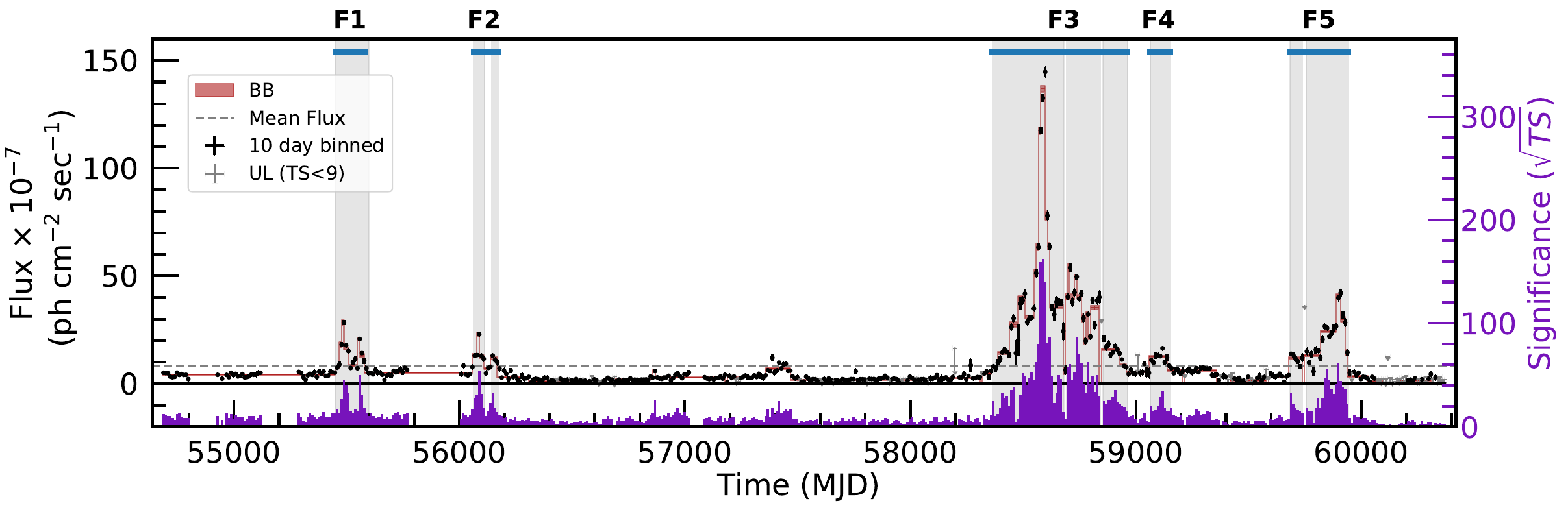}
\caption{ 10-day binned light curve of $\approx$ 15.5 years of Fermi-LAT observation of gravitationally lensed FSRQ PKS~1830$-$211. The grey region represents the high-flux state, and the white region represents the low-flux state. The light curve is divided into flaring epochs identified using HOP groups, marked by grey patches. HOP groups separated by less than 50 days are combined into flaring states, labeled as F1 to F5 (indicated by horizontal lines). The secondary y-axis (right) shows the detection significance as $\sqrt{\rm{TS}}$. Periods with TS $<$ 9 are represented by upper limits.}
\label{fig:fermi_LAT_10day_1830}
\end{figure*}

\section{Methods and Techniques}   \label{methods_techniques}
\subsection{Data Reduction - Fermi-LAT}
Despite being a lensed quasar, PKS~1830$-$211 appears as a point source due to the limited spatial resolution of the Fermi-LAT telescope. Regardless of this limitation, the Fermi-LAT telescope proves to be useful. It allows us to utilize the combined flux from the two anticipated lensed images, which have now coalesced into a signal exhibiting a distinct time delay. These delays are expected to provide valuable constraints on the emission size of the source. 

Fermi-LAT is a pair-conversion $\gamma$-ray telescope, sensitive to photons in the energy range of 20 MeV - 300 GeV \citep{Atwood_2009}. We selected data from 15.5 years of Fermi observations, covering the period from MJD 54683 to MJD 60373, within 10$\degr$  of the location of PKS~1830$-$211. This source, located approximately $\sim5\degr$ from the galactic center, is prone to galactic contamination. To minimize this, we selected photons with energies within 0.2 - 300 GeV.

To extract the photon statistics, the standard analysis procedure suggested by the Fermi Science Tools and the open-source Fermipy package \citep{Wood_2017} was used in the energy range between $0.2-300$ GeV, employing the latest instrument response function {\tt\string P8R3\_SOURCE\_V3}. A zenith angle cut of $90^\circ$, a {\tt\string GTMKTIME} cut of {\tt\string DATA\_QUAL} $>$ 0 \&\& {\tt\string LAT\_CONFIG==1}, and evtype=3 were used in the analysis. Only those events highly probable of being photons were considered for further analysis by applying a {\tt\string GTSELECT} cut on event class to account for {\tt\string SOURCE} class events using evclass=128.

A source model was prepared by including the source at RA = 278.413 and Dec = -21.075 and considering all the 4FGL catalog sources within $20\degr$ around the region of interest. The source is modelled with log-parabola model, parametrized as:
\begin{equation}  \label{eq:3}
\frac{dN}{dE}=N_\circ \left( \frac{E}{E_b} \right)^{-(\alpha + \beta(\log(E/E_b)))}
\end{equation}
where scale parameter $E_b$ was fixed to 4FGL catalog value of $645.56$\,MeV, $\alpha$ is specral index , $\beta$ is curvature parameter and $N_{o}$ is the Normalization.   Spectral parameters for the sources within $5^\circ$ of the region of interest were allowed to vary. Sources outside $5^\circ$ were fixed to 4FGL catalog values. Additionally, the background modeled with the diffuse galactic emission model ({\tt\string gll\_iem\_v07}) and the extragalactic isotropic diffuse emission model for point source analysis ({\tt\string iso\_P8R3\_SOURCE\_V3\_v1}) was allowed to vary.

We performed a binned likelihood analysis using {\tt\string GTLIKE} to evaluate the best-fit model parameters, including the source's spectrum and intensity at different epochs. The significance of detection is quantified using the Test Statistic (TS), defined as $\rm{TS}=-2 \ln(\mathcal{L}_0/\mathcal{L}_1)$, where $\mathcal{L}_0$ and $\mathcal{L}_1$ are the likelihood values without and with the point source at the position of interest, respectively. Only significant epochs with TS $>$ 9, predicted photons $>$ 3, and bins with flux greater than its uncertainty ($F_{t} > \sigma_{t}$) are considered for further analysis.

\subsection{Analysis Tools and Techniques}
\subsubsection{Bayesian Block and HOP algorithm}
To enable the detection and characterization of localized variability structures over time, we represent flux points and their uncertainties as step-functions using Bayesian Block \citep[BB;][]{Scargle_2013}. Each point of change in the block in the BB representation highlights a 3$\sigma$ variation from the previous block. The BB output is then processed by the HOP algorithm, which is based on a \textit{watershed} concept derived from topological data analysis \citep{Eisenstein1998ApJ}. HOP algorithm by itself identifies flaring states or periods of higher flux by clustering data points from neighboring regions where the flux exceeds a threshold. The combination of BB and HOP identifies a block higher than those before and after it as a peak. This approach then traces down from the peak in both directions, stopping when each subsequent block is lower than the previous one. Here, we use the mean flux as a lower threshold.

This technique divides the light curve into flaring and quiescent epochs, with consecutive connected BBs above the mean flux baseline referred to as a HOP group. The flare identification code developed by \citet{Wagner2021} differentiates between various HOP groups, leading to the identification of nine HOP groups, represented by grey patches for our source in Fig. \ref{fig:fermi_LAT_10day_1830}. The maximum time delay between the source and its lensed counterpart is expected to be approximately 70 days, as suggested by \citet{Barnacka2015ApJ}. Out of the nine HOP groups represented in Fig. \ref{fig:fermi_LAT_10day_1830}, some are less than 70 days apart. These close intervals suggest probable pairs of the source and its echo flare, arising from lensing. Therefore, we group HOP groups that are separated by less than 70 days, leading to five flaring states: F1 (MJD 55450 - 55600), F2 (MJD 56063, 56173), F3 (MJD 58363 - 58963), F4 (MJD 59063 - 59153), and F5 (MJD 59683 - 59943) as shown in Fig. \ref{fig:fermi_LAT_10day_1830}. The lag and magnification from these flaring groups are further discussed.

\subsubsection{Power Spectrum} \label{ps_PSRESP}
To identify the intrinsic temporal behavior of the time series, we evaluate the power spectral density (PSD) of high-energy $\gamma$-ray light curves. For a stochastic time series, the power distribution at each frequency typically follows a power-law (PL) ($P(f)\propto f^{-k}$) across various wavelengths and timescales, with an index ranging from approximately 1 to 3 \citep{Sobolewska2014ApJ, Finke2014ApJ, Nakagawa2013ApJ}. The average slope for the $\gamma$-ray PSD of the brightest 22 flat-spectrum radio quasars and 6 BL Lac objects is 1.5 and 1.7, respectively \citep{Abdo2010ApJ}. During the quiescent state, blazars typically exhibit temporal variability characterized by pink noise behavior, with $\alpha \sim 1$.

We compute the power-law variability index for the $1\,$ d and 12-hour binned Fermi-LAT light curve (LC) for flares F1 - F5 to quantify the temporal variability during the observed period, using the PSRESP method described in \cite{2014MNRAS.445..437M}, based on \cite{2002MNRAS.332..231U}. The obtained PSD is fitted with a PL model of the form PSD$(\nu)\propto\nu^{-k}$, where $k$ and $\nu$ are the spectral index and frequency, respectively. We simulate 1000 LCs with similar flux distribution and statistical variability as the observed LC using \cite{Connolly2015arXiv}. We have accounted for red noise leakage and aliasing effects as described in \cite{2022ApJ...927..214G}.


\begin{table*}
 \caption{PSD results} 
  \label{tab:psd_results}
  \begin{threeparttable}
    \begin{tabular}{cccccccccc}
    \hline
    Flux state\tnote{1} & Time period\tnote{2} & $T_{obs}$\tnote{3} & $N_{TS>9}/N_{tot}$\tnote{4} & $\Delta T_{min}$\tnote{5} & $\Delta T_{max}$\tnote{6} & $T_{mean}$\tnote{7} & $k \pm k_{err}$\tnote{8} & $p_{\beta}$\tnote{9} & $F_{var} \pm \Delta F_{var}$\tnote{10}\\
     & [day] & [day] &  & [day] & [day] & [day] &  &  & \\

    \hline
    \multirow{2}{*}{F1} & \multirow{2}{*}{MJD 55450 - 55600} & \multirow{2}{*}{$150$} & $123/150$ & $1$ & 5.0 & 1.22 & $0.91 \pm 0.39$ & 0.85 & 0.66$\pm$0.02\\
     &  &  & $168/360$ & $0.5$ & 5.0 & 0.77  & $0.96 \pm 0.24$ & 0.60 & 0.60$\pm$0.02 \\
    \multirow{2}{*}{F2} &\multirow{2}{*}{MJD 56063 - 56173} & \multirow{2}{*}{$110$}
    & $101/110$ & $1$ & 3.0 & 1.08 & $0.72 \pm 0.33$ & 0.05 & 0.46$\pm$0.03 \\
    
     &  &  & $153/220$ & $0.5$ & 4.0 & 0.71  & $1.15 \pm 0.24$ & 0.70 & 0.37$\pm$0.03 \\

    \multirow{2}{*}{F3} & \multirow{2}{*}{MJD 58363 - 58963} & \multirow{2}{*}{$600$} & $517/600$ & $1$ & 21.0  & 1.18 & $1.36 \pm 0.24$  & 0.83 & 0.88$\pm$0.01\\
    
     &  &  & $941/1200$ & $0.5$ & 21.0 & 0.64 & $1.37 \pm 0.22$ & 0.54 & 0.84$\pm$0.01\\
    \multirow{2}{*}{F4} & \multirow{2}{*}{MJD 59063 - 59153} & \multirow{2}{*}{$90$}  & $75/90$ & $1$ & 4.0 & 1.19  & $0.88 \pm 0.47$ & 0.14  & 0.19$\pm$0.05\\
     &  &  & $105/180$ & $0.5$ & 5.0 & 0.86 & $0.50 \pm 0.27$ & 0.34 & 0.12$\pm$0.08\\
    \multirow{2}{*}{F5} & \multirow{2}{*}{MJD 59683 - 59943} & \multirow{2}{*}{$260$} & $213/260$ & $1$ & 22.0 & 1.22  & 1.27$\pm$0.37 & 0.72 & 0.53$\pm$0.02\\
     &  &  & $355/320$ & $0.5$ & 21.5  & 0.73 & $1.23\pm0.20$ & 0.08 &  0.45$\pm$0.02\\

    \hline
    \end{tabular}
    \end{threeparttable}
    \begin{tablenotes}
    \small
 \item Note: (1)  Flux states extracted from use of BB and HOP algorithm (2) Period of the flaring states (3) Total exposure time (4) Fraction of points having TS $>$ 9 (5) Minimum sampling time in observed LC (6) Maximum sampling time in observed LC (7) Mean Sampling time  i.e. total observation time over a number of data points in that interval (8) The power law index for the power law model of PSD analysis using PSRESP method (9) p-value corresponding to the power law model. The power law model is considered a bad fit if $p_{\beta}\leq0.1$ as the rejection confidence for such model is $>90\%$ (10) Fractional variability
\end{tablenotes}
\end{table*}

\subsection{Estimating Time delay} \label{time_delay_estimation}

Gravitational lensing is often used as a promising tools for determining cosmological distances \citep{Refsdal1964MNRAS, Schechter1997ApJ, Blandford1992ARA&A}. Additionally, the time delay and magnification ratio derived at any wavelngth can explicate the location of the emission region relative to the central black hole \citep{Barnacka2014ApJ}. \citet{Atwood2007} predicted that LAT could detect delayed emission from bright lensed objects. High-energy observations of blazars exhibit significant variability due to the small emission region. The lensing-induced delay in photon arrival is expected to alter the intrinsic flux pattern of the source.  Unlike radio and optical telescopes, which can resolve magnified, multiple images of a lensed source, high-energy observations are often limited by poor spatial resolution. Consequently, the composite flux from source and its echo image appears as a point source to Fermi. This results in a repeated flux pattern spaced by \textit{a} days and demagnified by a factor of \textit{b} in the time domain. The total observed flux can be expressed as:

\begin{equation} \label{lensed_ts}
S_{\rm{obs}} = s(t) + s(t+a)/b    
\end{equation}

Thus, the total flux from the two images is integrated into the combined light curve when observed by high energy telescopes like Fermi. Thus, an added challenge is disentangling the repeated flares imprinted in the combined light curve from the apparent point source. \citet{Cheung2014ApJ} attempted to separate these flares and identified a leading and trailing component in the lensed blazar B0218+537.

In this work, we have used three techniques to estimate the lags in data using (1) autocorrelation Function, (2) double power spectrum, and (3) gaussian process regression. Fig. \ref{fig:fermi_LAT_10day_1830} represents the 5 flaring epochs that have been explored in further work.  

\begin{figure*}
\centering
\includegraphics [width=1.0\textwidth]{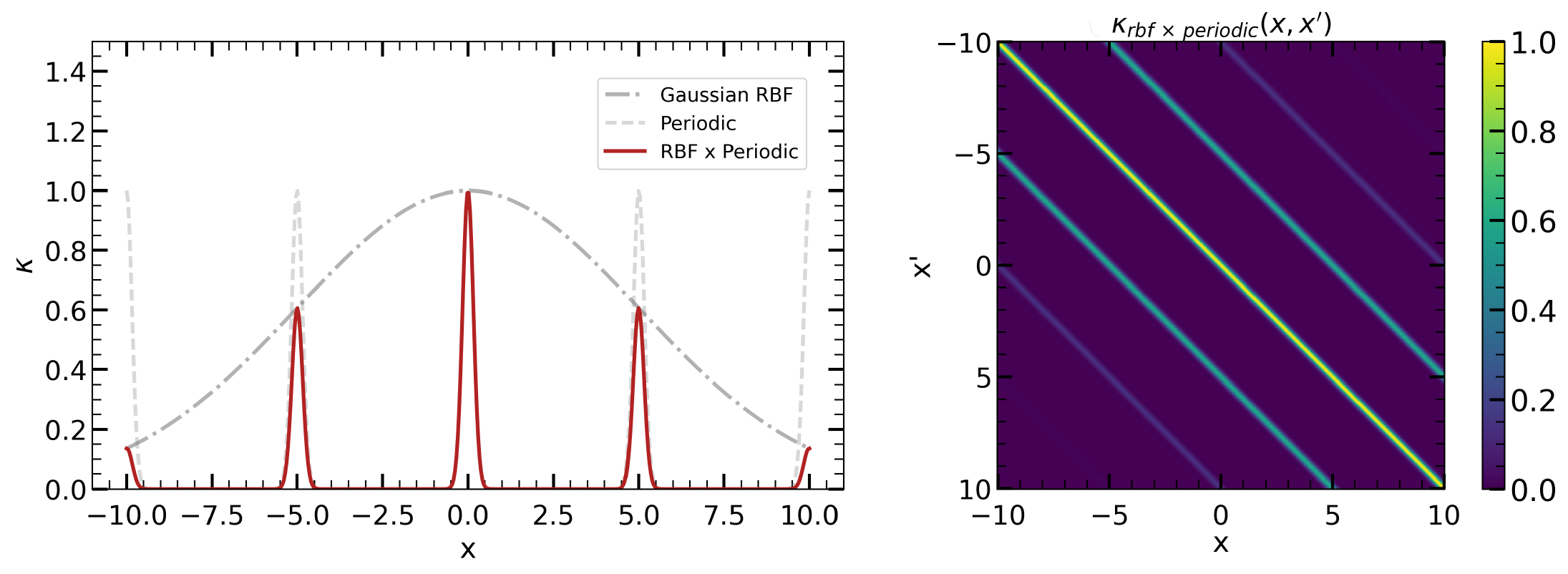}
\caption{(Left) Kernel visualization using covariance between each sample location and zeroth point for RBF, Periodic and $\rm{RBF}\times \rm{Periodic}$. (Right) Covariance matrix of the sample space for $\rm{RBF}\times \rm{Periodic}$ kernel where warmer colors indicate higher correlations.}

\label{fig:kernel}
\end{figure*}

\begin{figure}  
\centering
\includegraphics [width=0.5\textwidth]{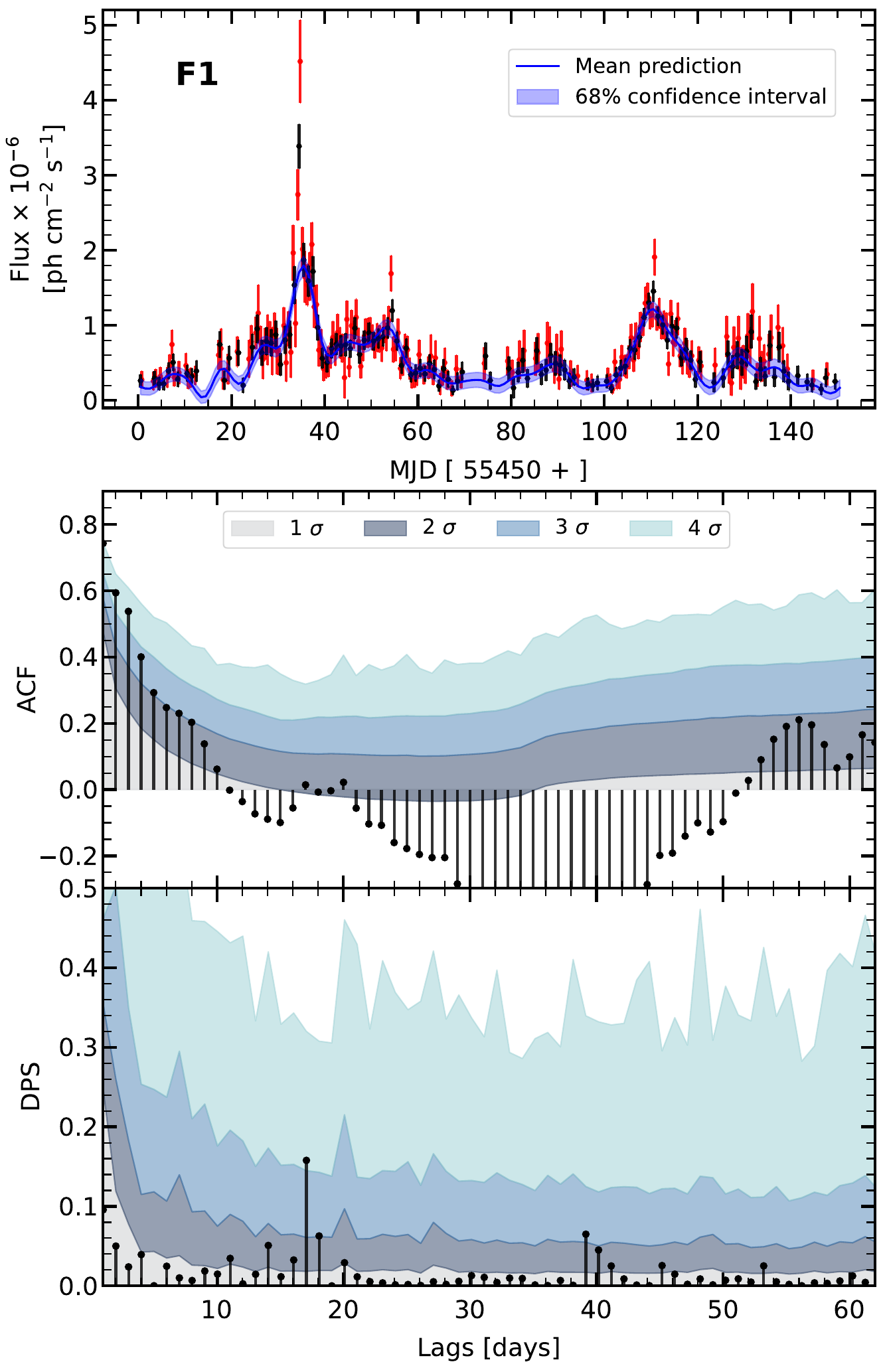}
\caption{ (Top panel) 1-day binned (black) and 12hr binned (red) light curve of flaring epochs F1 [MJD 55453 - 55583] (marked in Figure. \ref{fig:fermi_LAT_10day_1830}) of FSRQ PKS~1830$-$211. In blue is the  GPR predictions on 1 day binned data for the paramaters with largest marginal likelihood (Middle panel) ACF on 1-day binned light curve of F1 period (Bottom panel) DPS on 1-day binned light curve.}
\label{fig:LC_ACF_DPS_F1}
\end{figure}

\begin{figure}  
\centering
\includegraphics [width=0.5\textwidth]{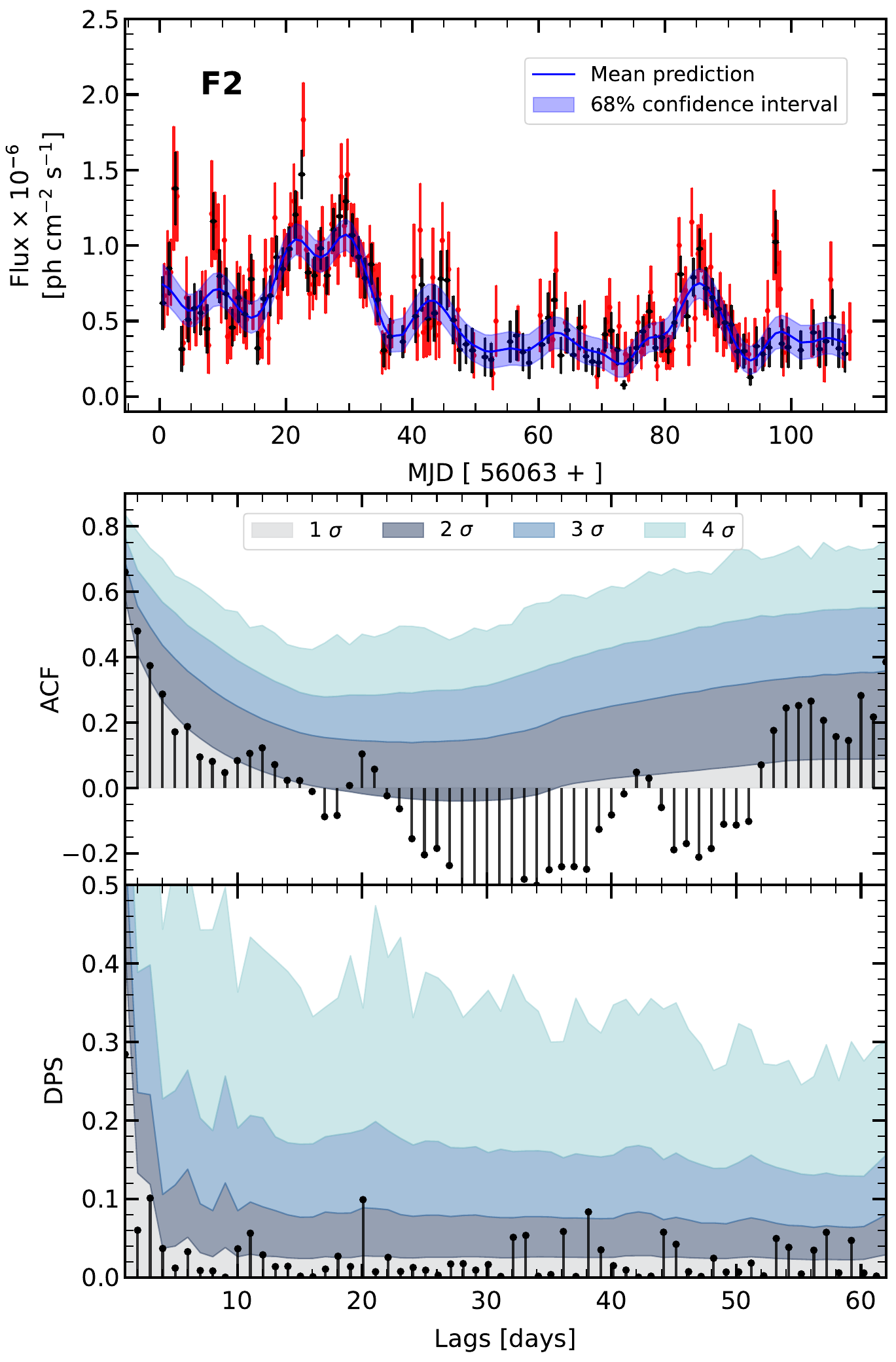}
\caption{ (Top panel) 1-day binned (black) and 12hr binned (red) light curve of flaring epochs F2 [MJD 56063 - 56173] (marked in Figure. \ref{fig:fermi_LAT_10day_1830}) of FSRQ PKS~1830$-$211. In blue is the  GPR predictions on 1-day binned data for the paramaters with largest marginal likelihood (Middle panel) ACF on 1-day binned light curve of F2 period (Bottom panel) DPS on 1-day binned light curve of F2 period }
\label{fig:LC_ACF_DPS_F2}
\end{figure}

\begin{figure}  
\centering
\includegraphics [width=0.5\textwidth]{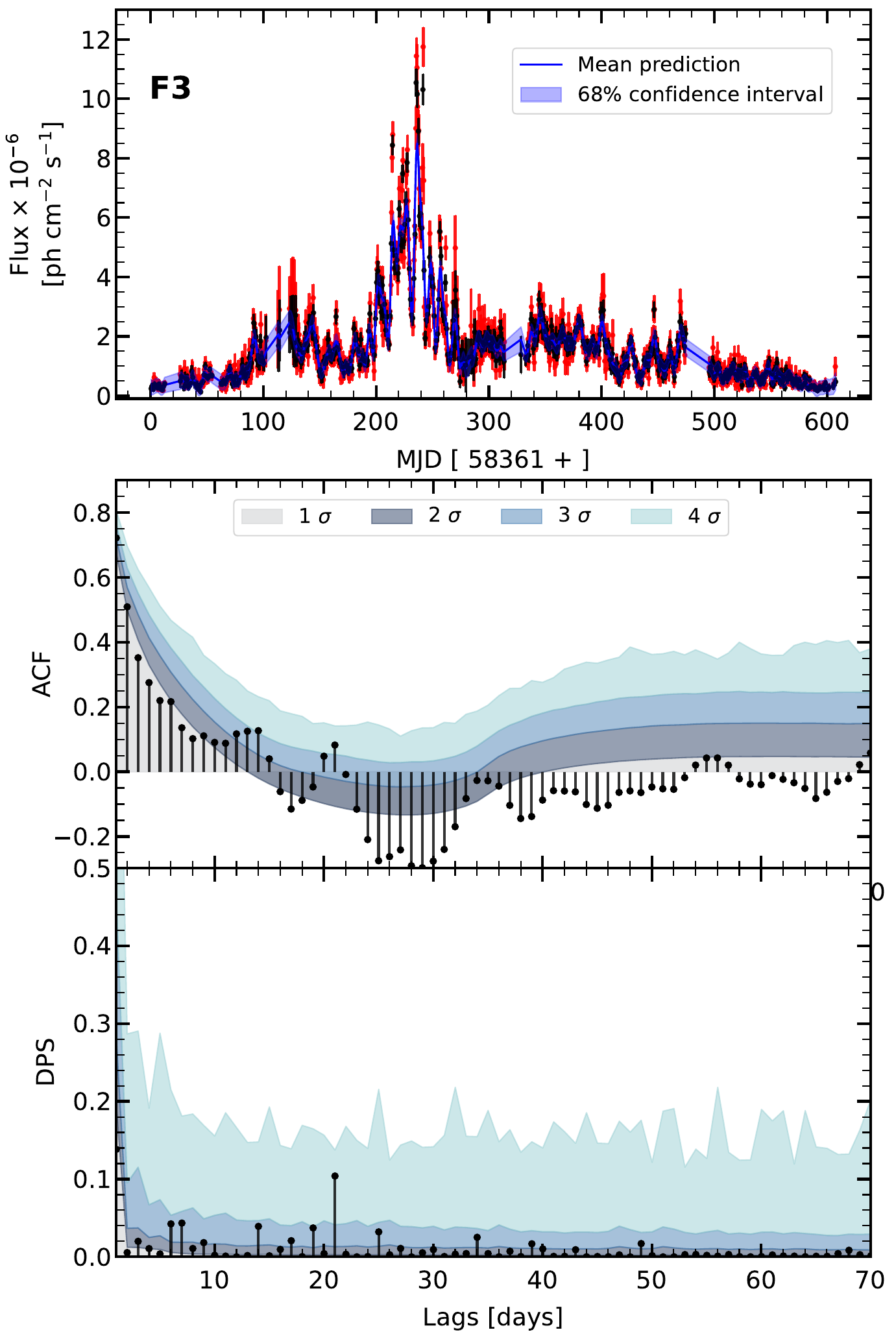}
\caption{ (Top panel) 1-day binned (black) and 12hr binned (red) light curve of flaring epochs F3 [MJD 58363 - 58963] (marked in Fig. \ref{fig:fermi_LAT_10day_1830}) of FSRQ PKS~1830$-$211. In blue is the  GPR predictions on 1-day binned data for the paramaters with largest marginal likelihood (Middle panel) ACF on 1-day binned light curve of F3 period (Bottom pane;) DPS on 1-day binned light curve of F3 period }
\label{fig:LC_ACF_DPS_F3}
\end{figure}

\begin{figure}  
\centering
\includegraphics [width=0.5\textwidth]{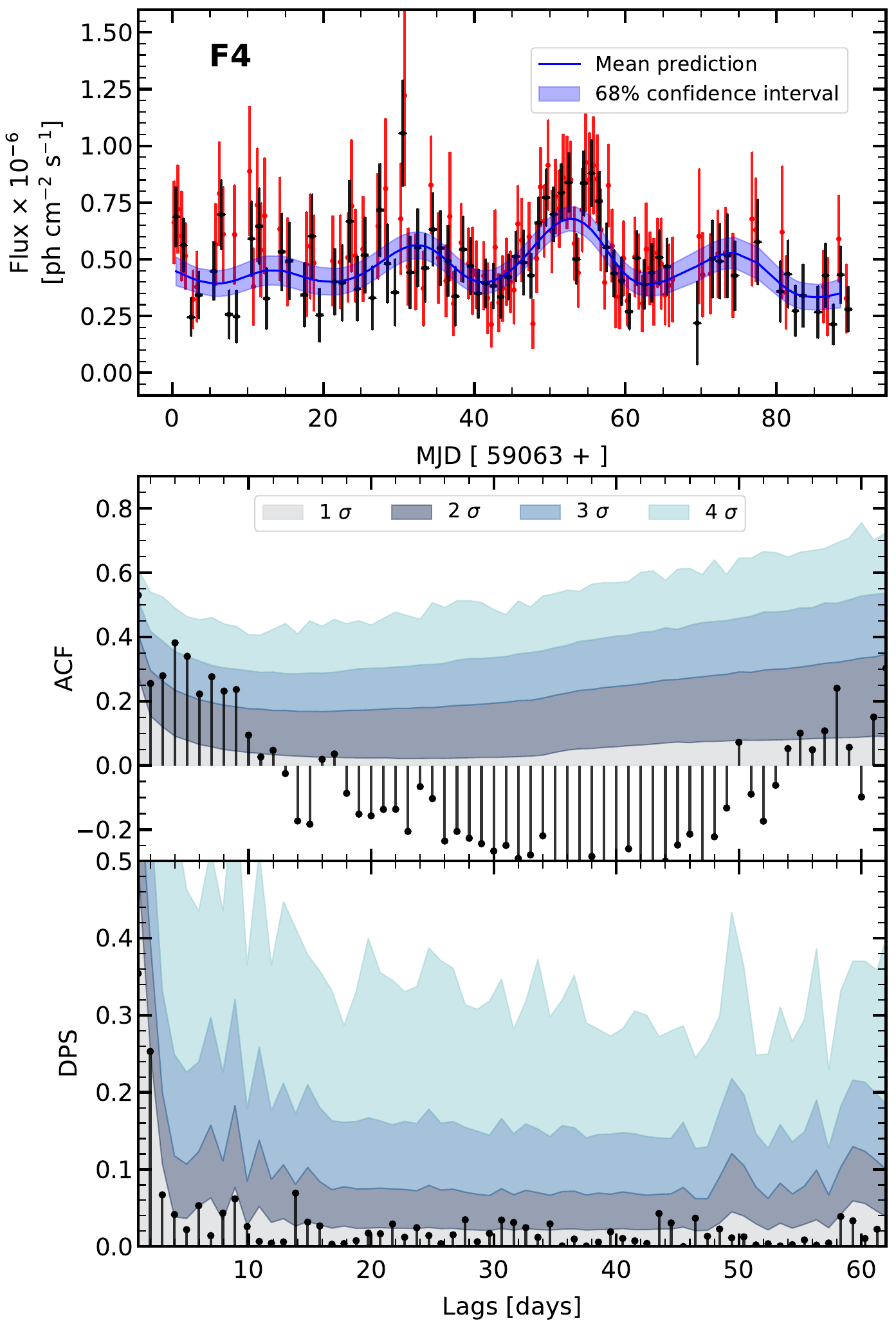}
\caption{ (Top panel) 1-day binned (black) and 12hr binned (red) light curve of flaring epochs F4 [MJD 59063 - 59153] (marked in Fig. \ref{fig:fermi_LAT_10day_1830}) of FSRQ PKS~1830$-$211. In blue is the  GPR predictions on 1-day binned data for the paramaters with largest marginal likelihood (Middle panel) ACF on 1-day binned light curve of F4 period (Bottom panel) DPS on 1-day binned light curve of F4 period }
\label{fig:LC_ACF_DPS_F4}
\end{figure}

\begin{figure}  
\centering
\includegraphics [width=0.5\textwidth]{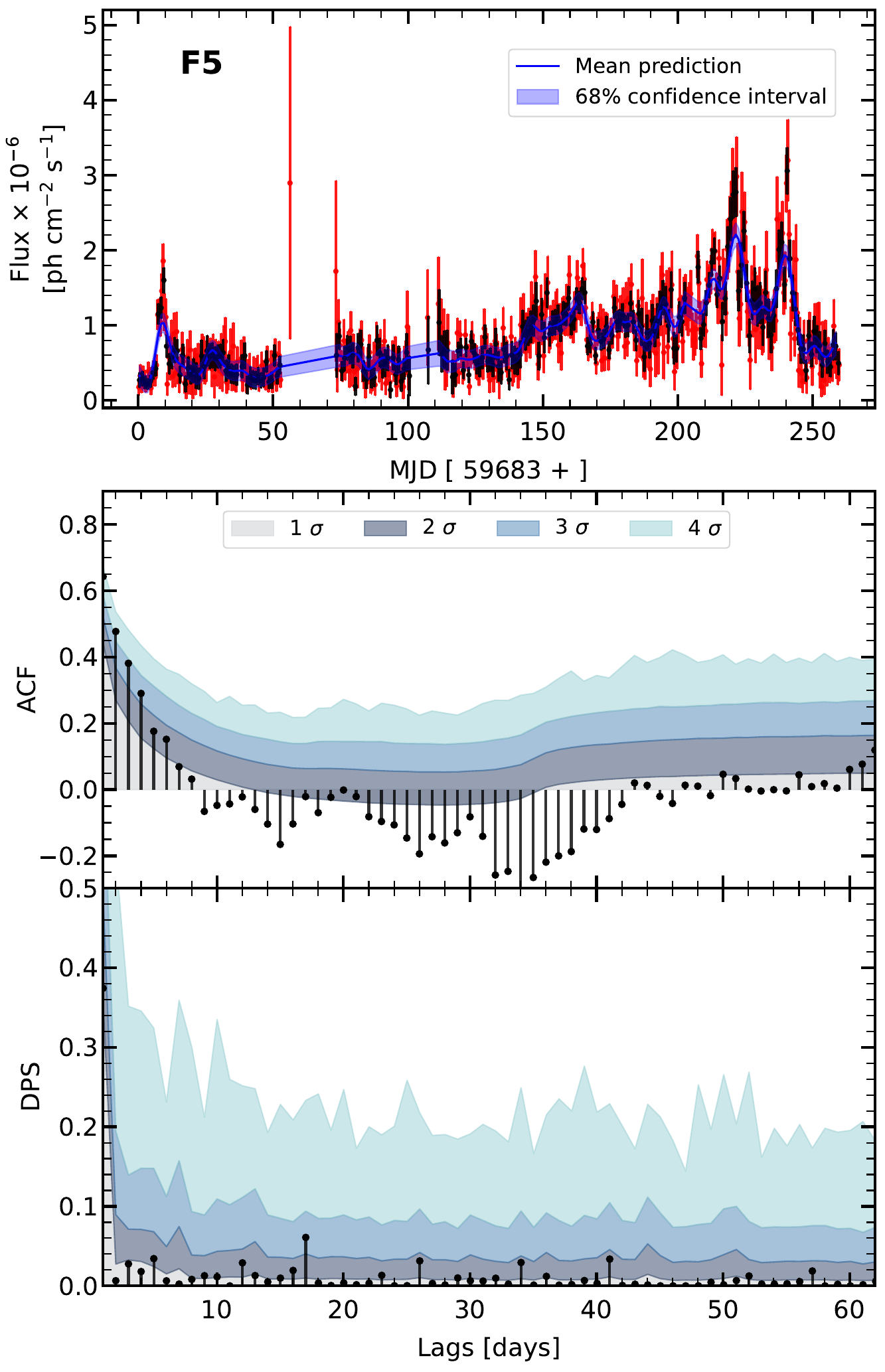}
\caption{ (Top panel) 1-day binned (black) and 12hr binned (red) light curve of flaring epochs F5 [MJD 59683 - 59943] (marked in Fig. \ref{fig:fermi_LAT_10day_1830}) of FSRQ PKS~1830$-$211. In blue is the  GPR predictions on 1-day binned data for the paramaters with largest marginal likelihood (Middle panel]) ACF on 1-day binned light curve of F5 period (Bottom panel) DPS on 1-day binned light curve of F5 period }
\label{fig:LC_ACF_DPS_F5}
\end{figure}

\begin{figure*}  
\centering
\includegraphics [width=1.0\textwidth]{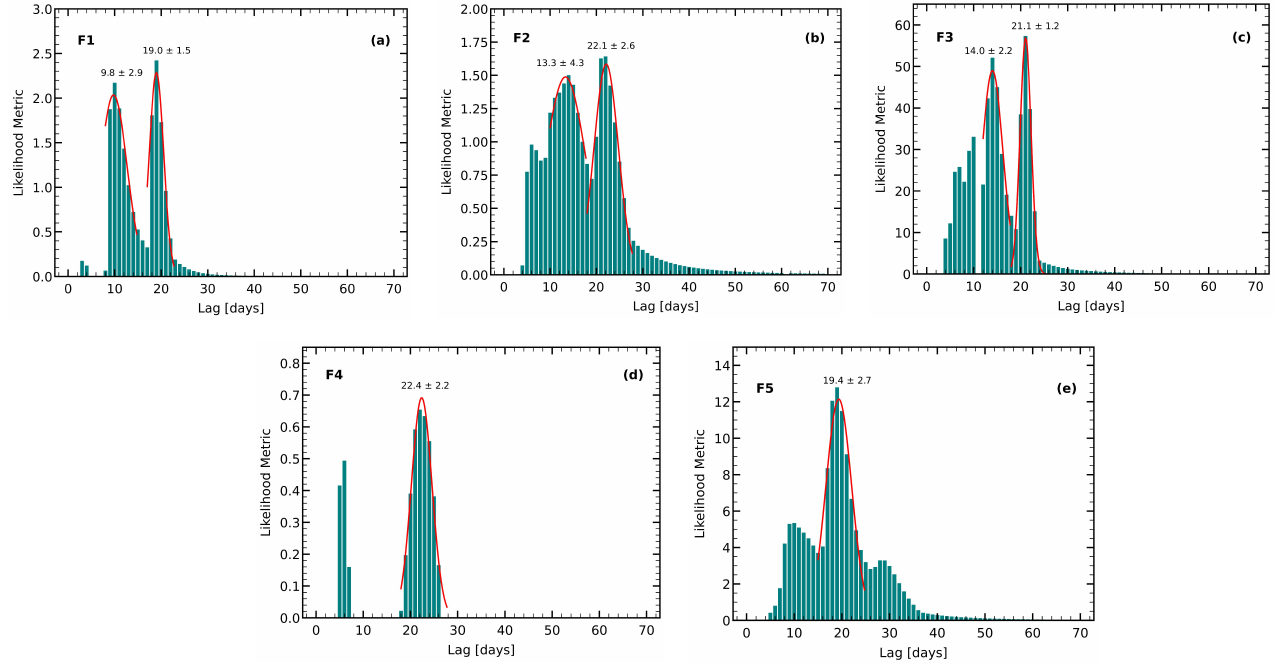}
\caption{ The likelihood metric for lags is derived using GPR. The bar represents the likelihood value for each lag, with the highest value indicating the most probable lag. The red Gaussian fit over the likelihood values represents the mean lag value estimated and its corresponding error bar.}
\label{fig:lag_likelihood}
\end{figure*}

\subsubsection{Auto-Correlation Function (ACF)}
The Autocorrelation function (ACF) is a standard statistical tool for assessing the similarity of a time series with a delayed copy of itself. ACF allows the identification of periodicity or repeated patterns in a signal, making it suitable for estimating lags in data.  

For a noise-dominated signal, variable structures are inherently present in power-law noise with an index greater than 1. \citet{Barnacka2015ApJ} described the role of ACF in deciphering the lag in noisy lensed signals and found that time delay detection is easier for a source with large variability index ($k$). Thus, a steep spectrum with spurious peaks improves the chances of confident detection.

The significance of the estimated lags is evaluated using the Monte Carlo simulation described in Section \ref{stat_significance}. To make the time series continuous, the epochs of no or less significant observation are interpolated with zeros as in \citet{Barnacka2015ApJ}.  We then performed autocorrelation on 1-day and 12-hour binned time series for flares F1 to F5. The results are discussed in Section \ref{results}. 

\subsubsection{Double Power Spectrum (DPS)}
A lensed time series is represented in equation \ref{lensed_ts}. Given the long, continuous, and evenly spaced nature of Fermi-LAT light curves, it is feasible to extract the lag from the data using the Fourier transform, as described by \citet{Barnacka2011A&A}, \citet{Barnacka2013arXiv}, and \citet{Barnacka2015ApJ}. This method was first used for lag estimation in lensed light curves by \citet{Barnacka2011A&A}. The idea is to take a Fourier transform of the first power spectrum of Equation \ref{lensed_ts}. The Fourier transform of the first component, $s(t)$, is $\Tilde{s}(f)$, and for the second component, it is $\Tilde{s}(f)e^{-2\pi i f a}$. Therefore, the observed signal in the frequency domain is transformed into:

\begin{equation}
\mathscr{F}(S_{\rm{obs})}= \Tilde{s}(f)(1+b^{-1}e^{-2\pi i f a})
\end{equation}

The first power spectrum (FPS) is the square modulus of Fourier transform of $S_{\rm{obs}}$, i.e. :

\begin{equation}
|\Tilde{S}(f)|^2= |\Tilde{s}(f)|^2 (1+b^{-2}+2b^{-1}cos{2\pi  f a})
\end{equation}

If an intrinsic lag is imprinted onto the signal, it should be encoded into the periodicity of the FPS with a time period inversely proportional to the lag. Therefore, a power spectrum of the FPS is expected to show a large amplitude of the time delay signal. \citet{Barnacka2011A&A} found that DPS is 90\% efficient in detecting the encoded lag in the signal, significantly improving over the 10

To correct for the smearing effect caused by the finite length of the signal and sampling in the data, the signal must undergo specific processing steps. We use the method described in \citet{Barnacka2015ApJ}, based on \citet{Brault1971A&A}, to extract the time delay present in the signal. This method can accurately extract time delays regardless of whether the light curve is white noise or red noise dominated and eliminates fake time delay peaks expected in red noise signals.

\subsubsection{Gaussian Process Regression (GPR)}
A Gaussian Process (GP) is a random process where any point $x$ in the real domain is a random variable $f(x)$, and the joint distribution of a finite number of these variables follows a Gaussian distribution. Mathematically, for a set of inputs $x_1$,$x_2$,…,$x_n$ with corresponding outputs $y_1$,$y_2$,…,$y_n$ , wherein y = f(x), the function values f(x) follow a joint Gaussian distribution. GP can be seen as a generalization of the infinite-dimensional multivariate Gaussian distribution. In a finite-dimensional Gaussian distribution, the correlation between random variables is defined using a covariance matrix. For an infinite-dimensional Gaussian distribution, this matrix is replaced by a "covariance function", known as a Kernel \citep{rasmussen2004_paper}. The mean function of the infinite-dimensional Gaussian is typically set to zero for easier computation, but the mean of the observational data is later added to obtain predictions on the original scale, leveraging the scaling property of Gaussian distributions. Standardization, which involves subtracting the mean and dividing by the standard deviation of the data before fitting the GP, is a common practice. 
\\
\textbf{Kernel selection:} The choice of kernel requires some prior information about the data. In our analysis, we incorporate the prior knowledge that the data exhibits a lag effect. This leads to the selection of the following kernel:

\begin{equation}
\kappa(x,x')=\exp\left(-\frac{|x - x'|^{2}}{2l^{2}}\right) \times \exp\left(-\frac{2}{l^{2}} \sin^2\left(\pi\frac{|x - x'|}{p}\right)\right)
\end{equation}

where $l$ is the length scale, and $p$ is the distance between repetitions. The first multiplicative element in this kernel corresponds to a Gaussian-shaped correlation function, while the second multiplicative element represents a periodic correlation.  Fig. \ref{fig:kernel} depicts the shape of the chosen kernel after multiplication. In this context, the periodicity parameter $p$ effectively functions as the lag.

\textbf{Hyper-parameter estimation :} The likelihood function can serve as an objective function for a non-linear optimization algorithm to obtain the maximum likelihood parameter values. In GPR, this is replaced by the log marginal likelihood, which incorporates both the data fit term and a penalty term to prevent overfitting. The log marginal likelihood consists of three terms added together:
The first term,  $-\frac{1}{2} \mathbf{y}^T (\mathbf{K}(\mathbf{X}, \mathbf{X'}) + \sigma_n^2 \mathbf{I})^{-1} \mathbf{y}$,  quantifies the quality of the fit. The second term,  $- \frac{1}{2} \log \det (\mathbf{K}(\mathbf{X}, \mathbf{X'}) + \sigma_n^2 \mathbf{I})$, helps avoid overfitting. The last term,  $- \frac{n}{2} \log(2\pi)$ is a normalization term to ensure a valid probability distribution, where $\mathbf{K(X,X')}$ is the covariance matrix, $\mathbf{I}$ is the identity matrix, and $n$ is the number of data points. 

We optimize the hyperparameters of the kernel $\mathbf{K}$. This study utilizes the scikit-learn GPR module \citep{scikit-learn} for easier computation. We optimize the length scale hyperparameter for various fixed periodicity hyperparameter values and obtain the log marginal likelihood profile across different lag values, as shown in Fig. \ref{fig:lag_likelihood}. We select a grid of lag values from 1 to 70 days in steps of 1 day. The upper limit for the lag is chosen based on prior information about the gravitationally lensed source \citep{Zhang2008ApJ}. Since the marginal likelihood can exhibit very similar values across different lags, we introduce a metric for better comparison, termed the "likelihood metric". Given that the marginal likelihood can be negative, we multiply it by -1 and subtract the maximum of this value from each marginal likelihood. The optimal lag then corresponds to the maximum value of the likelihood metric. Given the probabilistic nature of the method, the estimated lag is expected to exhibit a distribution centered around the true lag value. The uncertainties in the derived lag are quantified by analyzing the spread within this distribution.

\subsubsection{Statistical significance} \label{stat_significance}
The significance of spurious peaks in ACF and DPS must be assessed to determine whether the observed time delay is a result of chance or represents an intrinsic time delay within the signal.

We simulated $10^{5}$ light curves using the techniques described in \citet{10.1093/mnras/stt764} to generate artificial light curves with similar flux distribution and temporal variability as the observed light curve. This method allows for generating light curves with non-Gaussian distributions, overcoming a limitation of \citet{1995A&A...300..707T}. High-energy $\gamma$-ray light curves of blazars typically follow a log-normal flux distribution \citep{Romoli2018Galax, Bhatta2021ApJ}. As a result, the simulated light curves have identical statistical properties to the observed light curve.

To make the simulated light curves as realistic as possible, we included data gaps identical to the observed periods and interpolated them with zeros to ensure similar effects on the ACF and DPS of the simulated signal. We constructed cumulative probability distributions of the derived powers at each time delay. These distributions were then analyzed for 
$1\sigma$, $2\sigma$, $3\sigma$, $4\sigma$ chances of detection. Any significant ($>3\sigma$) powers corresponding to a time delay are considered as intrinsic time delays in the signal.

\section{Results} \label{results}
The HE light curve of PKS~1830-211 appears quite complex, with multiple flaring periods over the quiet states. Fig. \ref{fig:fermi_LAT_10day_1830} shows significant variability in the 10-day binned high-energy (200 MeV - 300 GeV) flux over time. This variability is evident from the fluctuating flux levels, with some periods displaying higher fractional variability than others (see Table \ref{tab:psd_results}). The flaring periods exhibit dominant pink noise behavior with a PSD power-law index of approximately 1. Notably, a transition from pink to red noise behavior is observed for the brightest flux state of the source, identified as F3 in this work (Table \ref{tab:psd_results}).
We use these flaring states to identify dominant emission zones, which should appear as twin pairs of flares separated by a specific time interval for a lensed blazar. The flaring epochs, which are above the mean flux levels, are identified using BBs, indicated by grey patches in Fig. \ref{fig:fermi_LAT_10day_1830}. Multiple blocks spaced less than 70 days apart are merged together, resulting in flaring states labeled F1, F2, F3, F4, and F5.

\begin{table}
    \centering
    \begin{tabular}{ccc}
    \hline
        Flare state & $\alpha \pm \delta \alpha$ & $\beta \pm \delta \beta$\\
        \hline
        F1 & 2.39 $\pm$ 0.03 & 0.16 $\pm$ 0.02 \\
        F2 & 2.29 $\pm$ 0.03 & 0.10 $\pm$ 0.02\\
        F3 & 2.38 $\pm$ 0.01 & 0.15 $\pm$ 0.01 \\
        F4 & 2.55 $\pm$ 0.04 & 0.13 $\pm$ 0.04\\
        F5 & 2.41 $\pm$ 0.02  & 0.14 $\pm$ 0.02\\
        Quiet state &  2.47 $\pm$ 0.02 & 0.08 $\pm$ 0.01 \\
        \hline
    \end{tabular}
    \caption{Fermi-LAT Spectral parameters for the chosen flaring states and the quiet state. The parameters are derived from the fitted log-parabola model. }
    \label{tab:states_spectral_params}
\end{table}

The flaring periods, except for F4, exhibit similar $\alpha$ parameters in HE Fermi-LAT spectrum, indicating a consistent physical process within a $3\sigma$ range. Additionally, the $\beta$ values for all periods are consistent, suggesting a similar influence of external seed photons in the production of high-energy $\gamma$-rays. The spectral parameters for the flaring periods F1-F5 are summarized in Table \ref{tab:states_spectral_params}.

The time lag between these counterpart flares from the lensed images is estimated using the three methods described in Section \ref{time_delay_estimation}. The maximum time delay between the lensed images is given by $\sim 6 \left( \frac{z_g}{0.1} \right) \left( 2h \right)^{-1}$ days, where $z_g$ is the redshift of the lensing galaxy and $h$ is the Hubble constant in units of 100 $\rm{km\,s^{-1}\,Mpc^{-1}}$ \citep{Zhang2008ApJ}. Using a redshift of $z_g=0.89$ and $h=75, \rm{km\,s^{-1}\,Mpc^{-1}}$, the maximum time lag is approximately 71 days. This represents the maximum time delay between the mirage images when the source is near the Einstein ring. For large delays, the magnification ratio is expected to be significant. Thus, detecting a large delay is unlikely since the trailing component would be demagnified beyond the sensitivity of Fermi-LAT. Additionally, to explore the full range of potential time delays, the length of the lightcurve must be at least twice the longest time delay.

Detecting the trailing counterpart during a quiescent state of a blazar is challenging due to its expected demagnification, which impedes detection. However, the sensitivity of Fermi-LAT allows for the detection of multiple flaring states (F1 to F5). Dominant pink noise during flaring epochs creates spurious peaks, increasing the chances of detecting both trailing and leading components. The source shows the largest variability for F3 and F1, with respective fractional variability values  \citep[F$_{var}$;][]{Vaughan2003MNRAS} of $0.88\pm0.01$ and $0.66\pm0.02$ (Table \ref{tab:psd_results}). Our objective is to identify time delays from flares with sufficient magnification to detect both leading and trailing components. Whenever possible, pairs of leading and trailing flares, spaced within the identified time lag, are selected to study the spectral properties of source and echo flare and draw the relationship between lag and magnification. To analyze the flare variability properties, we fitted exponential flares to sharp, distinct features in the light curve using the form:

\begin{equation}   \label{eq:6}
F(t)= F_{0} \times \left[ \exp \left( \frac{t_o-t}{\tau_{\rm{rise}}}\right) + \exp \left( \frac{t-t_o}{\tau_{\rm{decay}}}\right) \right]
\end{equation}

Here, $\tau_{\rm{rise}}$ and $\tau_{\rm{decay}}$ represent the rise and decay timescales, respectively; $t_o$ is the peak time, and $F_0$ represents half of the peak flux of the flare at time $t_o$.
 
\begin{table}
    \centering
    \begin{tabular}{cccc}
    \hline
        Flare & $\rm{Lag_{ACF}}$ & $\rm{Lag_{DPS}}$ & $\rm{Lag_{GPR}}$\\
        \hline
        F1 & - & 17$\pm$1.5 ($>3\sigma$) & 19.0$\pm$1.5\\
        \hline
        F2 & 20.3$\pm$2.3 & 20.0$\pm$0.5 ($\sim 2\sigma$) & 22.1$\pm$2.6 \\
        \hline
        F3 & 20.5$\pm$1.0 & 21.0$\,\pm\,$0.5 ($\sim 3\sigma$) & 21.1$\pm$1.2\\
        \hline
        F4 & - & 14.0$\pm$0.5 ($< 2\sigma$) & 22.4$\pm$2.2 \\
        \hline
        F5 & - & 17.0$\pm$0.5 ($>2\sigma$) & 19.4$\pm$2.7\\
        
        \hline
    \end{tabular}
    \caption{Estimated lags for Flares F1, F2, F3, F4, F5 using the three methods (a) Autocorrelation Function (b) Double Power Spectrum (c) Gaussian Process regression}
    \label{tab:time_delay}
\end{table}

\subsection{Flare 1 - MJD 55453 - 55583}
The 1-day and 12-hour binned light curve of flare F1 is shown in Fig. \ref{fig:LC_ACF_DPS_F1}(a). The flaring period spans 150 days, with 18\% of the flux points not resulting in significant detection. The flare exhibits a sharp peak from MJD 55483 to MJD 55488 and another distinct feature from MJD 55553 to MJD 55573 in the $\gamma$-ray light curve.

The light curve exhibits prominent pink noise behavior with a power-law index of $k = 0.91 \pm 0.39$. Monte Carlo simulations were conducted to quantify the significance of the time delay by generating light curves with similar power spectral density indices. The Autocorrelation Function did not result in a significant detection, but a feature emerged at $55 \pm 2$ days ($\sim 2\sigma$), likely an artifact of the Fermi light curve due to the spacecraft's precession period of $53.4$ days (Fig. \ref{fig:LC_ACF_DPS_F1}(b)).

The DPS method is more prone to detecting spurious time delays. The DPS on a 1-day binned light curve peaked at $17 \pm 1.5$ days with above $3\sigma$ significance as shown in Fig. \ref{fig:LC_ACF_DPS_F1}(c). Similarly, GPR analysis on the 1-day binned light curve identified the maximum marginal likelihood metric at $19.0 \pm 1.5$ days. The corresponding best-fit GPR light curve is shown in Fig. \ref{fig:lag_likelihood}(a). The marginal likelihood peaking at $9.8 \pm 2.9$ days could represent a lower harmonic of the $19.0 \pm 1.5$ day delay, which aligns with the periodic nature of the selected kernel. Consistent results have been reported in previous studies, such as a lag of $19 \pm 1$ days by \citet{Abdo2015ApJ} and $17.9 \pm 7.1$ days by \citet{Barnacka2015ApJ} using ACF.

The maximum peak in the light curve at MJD 55483 to MJD 55488 and the resulting echo flare (F11 in Fig. \ref{fig:flare_expon_fit}(a)) after 20.0$\pm$1.1 days can be a proxy for evaluating the magnification ratio, resulting in a ratio of $\sim$2.8. Another distinct feature at MJD 55553 to MJD 55573 (F12), considering a similar time delay, shows a magnification ratio of  $\sim$1.6 as in Fig. \ref{fig:flare_expon_fit}(a).

\subsection{Flare 2 - MJD 56063 - 56173}

Fig. \ref{fig:LC_ACF_DPS_F2}(a) shows the 1-day and 12-hour binned light curves. This epoch is above the mean flux level, with periods before and after significantly below the mean. The 110-day-long light curve follows a pink noise power-law time series with an index of $k=0.72 \pm 0.33$. Simulated light curves with similar indices were generated to measure the significance of the lag in the signal.

The autocorrelation on F2 in Fig. \ref{fig:LC_ACF_DPS_F2}(b) shows two features with significance close to $2\sigma$: $11.0 \pm 2.3$ days and $20.3 \pm 0.5$ days, consistent with \citet{Barnacka2015ApJ}. An additional lag close to $2\sigma$ at $55.7 \pm 2.2$ days appears in both the 1-day and 12-hour binned light curves, possibly an artifact of the Fermi telescope's processing period, similar to Flare F1.

The DPS method on the 1-day binned light curve detected a time delay of $20 \pm 0.5$ days with more than $2\sigma$ significance (Fig. \ref{fig:LC_ACF_DPS_F2}(c)). Consistent lags are found using GPR, with increased marginal likelihood metrics at $13.3 \pm 4.3$ and $22.1 \pm 2.6$ days, aligning with the ACF results. The likelihood distribution of GPR on Flare F2 is as in Fig. \ref{fig:lag_likelihood}(b). The $13.3 \pm 4.3$ day lag in GPR could be a lower harmonic of the $22.1$ day delay in the light curve.

Disentangling the light curve to identify flares and their echo flare image is challenging for Flare F2. The double-peak structure from MJD 56081 to MJD 56098 has a demagnified delayed counterpart appearing from MJD 56101 to MJD 56118, with a demagnification of approximately 1.9 as shown in Fig. \ref{fig:flare_expon_fit}(b)). No echo counterpart to the broad feature from MJD 56142 to MJD 56156 is visible within the 20-day period. This absence suggests a much larger demagnification corresponding to longer time delays.

\subsection{Flare 3 - MJD 58363 - 58963}
The 1-day and 12-hour binned light curve for Flare 3 is shown in Fig. \ref{fig:LC_ACF_DPS_F3}(a). This flare represents the brightest flux state of the source, with the flux reaching 14 times its average level. Notably, it is also the longest-lasting flare analyzed in this work, spanning a total of 600 days. The flare exhibits the highest power spectral index, with characteristic noise behavior between pink and red noise types. Multiple flares appear to be superimposed on an underlying envelope, as illustrated in Fig. \ref{fig:LC_ACF_DPS_F3}(a).

The ACF, as displayed in Fig. \ref{fig:LC_ACF_DPS_F3}(b)) reveals two features with significance greater than $3\sigma$: one at $12 \pm 1.8$ days and another at $21.1 \pm 1.2$ days. For comparison, the DPS method calculates a prominent lag at $21.0 \pm 0.5$ days ($\gtrsim 3.5\sigma$) and $19.0 \pm 0.5$ days ($=3\sigma$). Less significant detections, but still above $\gtrsim 2.5\sigma$, are found at $14 \pm 0.5$ days and $25 \pm 0.5$ days. This suggests multiple lag values imprinted on the flares. The estimated lag values could also be a reflection of the time difference between subsets of flares. 

A consistent lag detected by GPR at $14.0 \pm 2.2$ and $21.1 \pm 1.2$ days is simultaneously observed by ACF and DPS (Figure \ref{fig:lag_likelihood}(c)).

Due to the dense overlap of flaring periods, identifying associated leading and trailing counterparts is extremely challenging during such a flaring period. Therefore, estimating magnification by flare identification is not performed for Flare 3.

\subsection{Flare 4 - MJD 59063 - 59153}
Fig. \ref{fig:LC_ACF_DPS_F4} shows the 1-day and 12-hour binned light curve for Flare 4, which spans 90 days and is the least bright of the five states analyzed. The power spectral density for the 1-day binned data yields $k = 0.88 \pm 0.47$. Ideally, the sample length should be twice the maximum expected time delay of 70 days, so the 90-day light curves reduce the likelihood of detecting time delays. Additionally, F4 has the least variability in lightcurve with $F_{var,F4}=0.19\pm0.05$ days. Such small fractional variability highlights the absence of significant variable points in the light curve.

No significant time delay is detected by ACF and DPS in the 1-day binned data. GPR estimates a lag of $22.4 \pm 2.2$ days as shown in Fig. \ref{fig:lag_likelihood}(d). The absence of significant echo flares for the flux rise at MJD 59113 - 59123 (Fig. \ref{fig:LC_ACF_DPS_F4}(a)) makes it difficult to conclude if a lensed image is detectable within the telescope's sensitivity.

\subsection{Flare 5 - MJD 59683 - 59943}
Fig. \ref{fig:LC_ACF_DPS_F5}(a) shows the 1-day and 12-hour binned light curve for Flare 5, with multiple peaks visible over the 260-day period. At least 18.1\% of the data contains gaps or periods with significantly low detection, which we interpolated with zeros. No significant time lags were found using ACF. However, DPS estimates a $17 \pm 0.5$ day time lag with more than $2\sigma$ significance, consistent with the predicted time delay from GPR at $19.4 \pm 2.7$ days (See Fig. \ref{fig:lag_likelihood}(e)). 

\begin{figure}  
\centering
\includegraphics [width=0.5\textwidth]{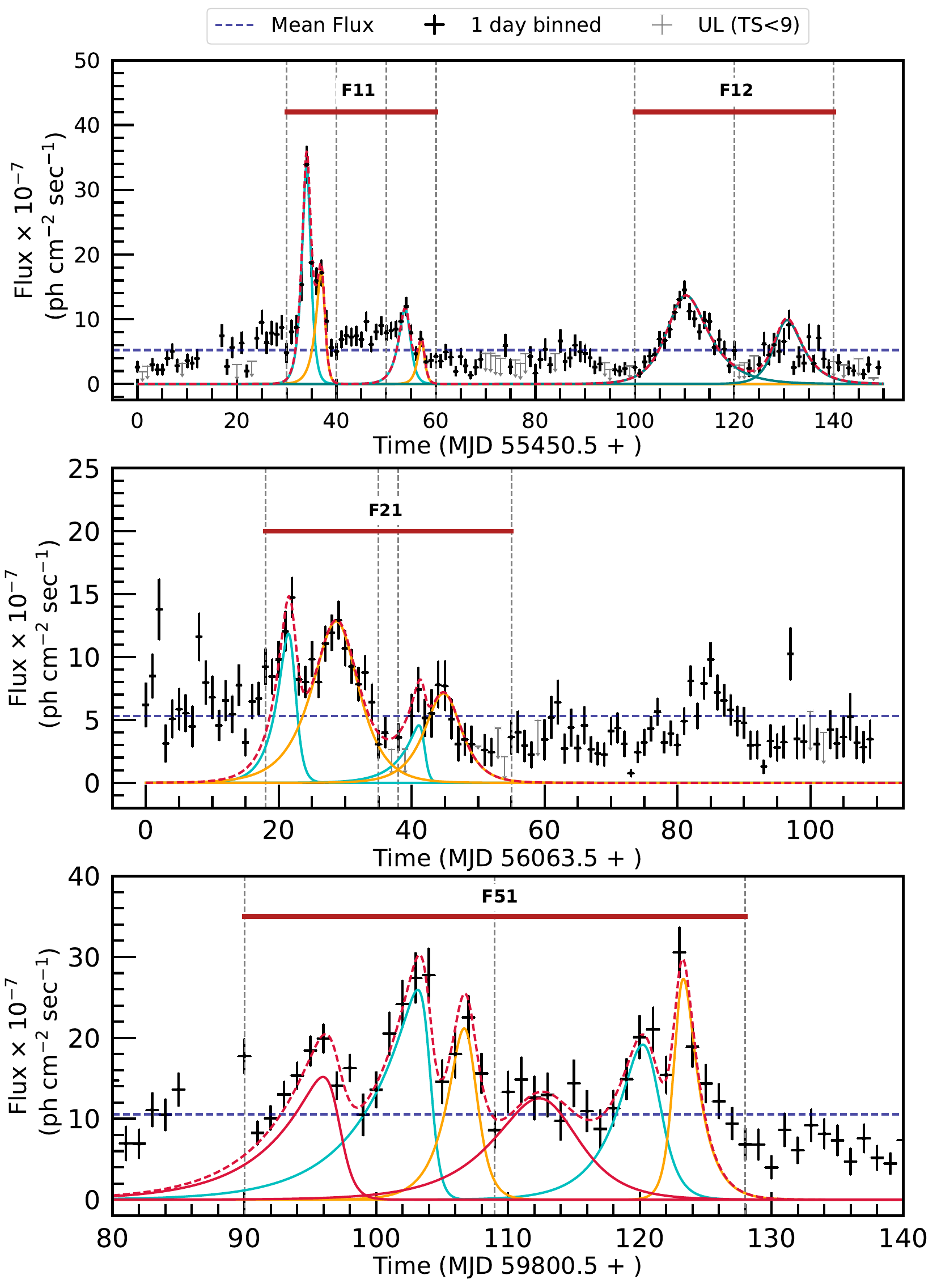}
\caption{High flux states of Flare F1, F2 and zoomed section of F5 (MJD 59880 - 59940) and fitted exponential flare using equation \ref{eq:6} . The vertical lines represent the source and echo pair for the lensed flares. The exponential fits with similar colors are considered possible pairs of source and echo flares.}
\label{fig:flare_expon_fit}
\end{figure}

The bright peaks at MJD 59890 to MJD 59928 (F51) were fitted with exponential function as shown in Fig. \ref{fig:flare_expon_fit}(c). However, due to the emergence of multiple overlapping flares, associating flares to identify magnification is challenging.

\begin{figure}  
\centering
\includegraphics [width=0.5\textwidth]{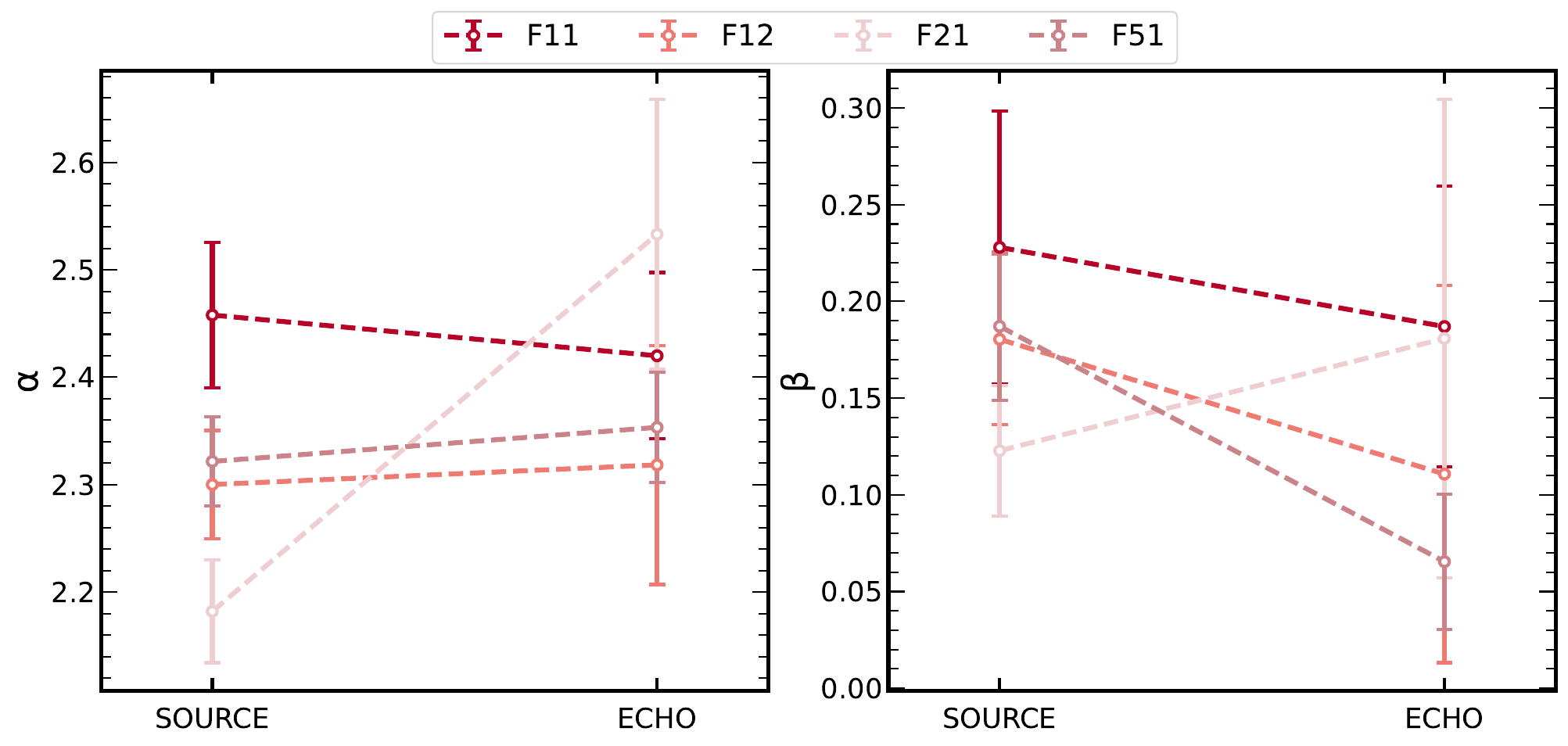}
\caption{High energy spectral parameter of the flare and its associated echo flare. (left) $\alpha$ and (right) $\beta$ for the fitted log-parabola model.}
\label{fig:flare_index}
\end{figure}

\section{Discussion}  \label{discussion}
The time delay between two lensed images of a quasar, such as PKS 1830$-$211, is typically evident from systematic changes in flux density in the radio band for the leading and trailing components. However, in the $\gamma$-ray band, the resolution of high-energy telescopes constrains the detection of resolved images. As a result, the combined flux evolution from the two lensed images appears merged into a single point source.

The time delay observed from the merged images reflects the distribution of emission sites on the lens plane around the mass-weighted center of the lens \citep{Barnacka2014ApJ}. Time delays, along with magnifications, are fundamental for localizing the emitting region relative to the black hole. Typically, a smaller time delay is associated with areas close to the base of the jet. Emissions originating near the base of the jets are typically reflected as the fast variability in high-energy $\gamma$-ray light curves.

In this work, we employed three methods to investigate the observed time lag in the stochastic time series. The resulting lags are listed in Table \ref{tab:time_delay}. We devised a novel technique, Gaussian Process Regression, to extract the lag in the signal. From our analysis, we found that the autocorrelation function is unable to efficiently extract the intrinsic time delay present in most signals, partially due to its noise dependence. However, among the five flares studied, autocorrelation successfully detected the intrinsic lag in the time series for Flare 3 with a significance of more than $3\sigma$. Flare 3 exhibited noise behavior between pink and red, and the emergence of multiple flares resulted in significant detection. Similar results were obtained using the Double Power Spectrum and Gaussian Process Regression.

Our results suggest the presence of a consistent time delay of approximately 20 days during the flaring state of the source, as determined by the three methods used in this work. This indicates a similar orientation of the emitting site around the mass distribution of the lens. Consequently, $\gamma$-ray emission consistently occurs in similar regions of the jet across all flaring states.
The inferred time delay aligns with the estimated lag reported in \citep{Barnacka2015ApJ} during the high states, indicating that the origin of the $\gamma$-rays is likely within the core. The detection of rapid variability, with $t_{var} \sim 0.38 \pm 0.22$ days, implies an emission region size of $r_{emm} = c \delta t_{var}/(1+z) = 2.8 \times 10^{15}$ cm at a distance of $R_{diss} = 2c\Gamma^2 t_{var} = 0.064$ pc, thereby confirming that the high-energy emission is localized within the core on sub-parsec scales.

\citet{Lovell1996ApJ} reported a lag of $26_{-5}^{+4}$ days using the Australian Telescope Compact Array at 8.6 GHz. Similar time delays were estimated in the $\gamma$-ray band during the quiet state of the source \citep{Barnacka2011A&A}. The lag observed during high states in this study suggests a different origin for flaring $\gamma$-ray emission. The inconsistency between $\gamma$-ray and radio lags indicates different dissipation sites for these emissions, especially during the source's high state. Radio emissions are typically expected from the outer regions of the parsec-scale jet. A small, compact jet leads to synchrotron self-absorption, making radio emission unlikely in the inner parsec scales. Shorter time delays during active states imply that $\gamma$-ray dissipation occurs closer to the central engine, whereas radio dissipation occurs farther out in the jet. This has been observed as the absorption of high-energy photons with energies greater than 10 GeV during high states under the influence of BLR photons at sub-parsec scale jet \citep{Agarwal2024ApJ}.

The high-energy spectral properties of the source and the echo flare are consistent within 3$\sigma$ (See Fig. \ref{fig:flare_index}). A change in the spectral properties would imply a difference in the influence of soft seed photons on $\gamma-$ray photon through $\gamma-\gamma$ absorption on passing through a more luminous region of the lensing galaxy. A $\sim2.8\sigma$ deviation was observed in the spectral index for flare F21. The identical beta parameters for the four possible lensed flares suggest a similar influence of external seed photons from the local jet environment, the EBL, and the intervening galaxy on the high-energy spectrum. Since absorption affects all of the flares in the same way, they originated from similar regions of the jet. Our analysis focused on flares with a clear source and a demagnified lensed echo flare at an average flux level, leading us to select flares F1, F2, and F5. Due to the presence of multiple overlapping flares, identifying individual flares and their echoes for flare F3 was inefficient, likely due to the merging of multiple flares. Exponential fitting on individual flares reveals a linear relationship between lag and magnification. However, further studies are needed to identify more clean flares. This suggests that a smaller emission region is confined close to the base of the jet, while a larger magnification implies a larger emission region located farther out in the jet.

\section{Summary}    \label{summary}
Strong gravitational lensing in $\gamma$-ray bright blazars can identify the locations of $\gamma$-ray dissipation during both quiescent and active states. The variation in time delays observed during periods of active $\gamma$-ray flux suggests different emission regions within the jet compared to those during low flux states \citep{Barnacka2011A&A}. The consistent lag across five flaring states indicates a similar origin for the high-energy $\gamma$-ray activity within the radio core. This contrasts with the larger lag observed during quieter $\gamma$-ray periods and the consistent time delays from radio observations, which suggest that such emission occurs farther from the central engine than that during flaring periods. Such time delays caused by gravitational lensing of a background source by a foreground object could help constrain the Hubble parameter \citep{Refsdal1964MNRAS}.

We introduce a novel technique for estimating time delays in long, continuous light curves from Fermi-LAT. Detecting these time delays could be crucial in identifying hidden lensed blazars during flaring periods that are not recognized as lensed sources in radio wavelengths. The signatures of such time delays could provide insights into distant blazars and previously unidentified $\gamma$-ray sources. Future surveys, including those conducted by the SKA, will likely discover many such lensed quasars. Extensive multi-wavelength searches of these systems could offer valuable insights into the origins of radiation and provide a magnified view limited by current telescopes.

\section{Acknowledgements}
We thank the refree, Dr. Nachiketa Chakraborty, for constructive feedback, which has helped improve the manuscript. SA and AS acknowledges Dr. Bhargav Vaidya for useful discussions on the work which helped improve the manuscript. AS acknowledge support for computational facility from DST-SERB grant CRG/2022/009332. This research work has made use of archival data, software and web tools obtained from NASA’s High Energy Astrophysics Science Archive Research Center (HEASARC) and Fermi gamma-ray telescope Support centre, a service of the Goddard Space Flight Center and the Smithsonian Astrophysical Observatory. 

\section*{Data Availability}
The data will be made available upon reasonable request.


\bibliographystyle{mnras}
\bibliography{main} 

\bsp	
\label{lastpage}
\end{document}